# Classroom as a Microcosm:
# Teaching Culturally Diverse Students

**Senad Bećirović, PhD**     **Damir Bešlija**

International Burch University

Sarajevo, Bosnia and Herzegovina

*senad.becirovic@ibu.edu.ba*     *d.beslija@gmail.com*

*Abstract: The twenty-first century is the century of encounter of the different races, nations, cultures, religions and customs. In the twenty-first century, man is more and more exposed to various influences that leave a trace on the entire sphere of his social life, including education. Given that education systems play one of the key roles in the formation of both physically and morally healthy communities, it is of an enormous importance to analyze the phenomenon of a classroom composed of culturally diverse students. Each individual is nowadays exposed to various influences that leave a trace on the educational sphere of his social life. Taken into consideration how educational institutions have become more and more diverse in terms of cultures, views and perspectives it is of a great importance to analyze the phenomenon of a multicultural education. Moreover, it is of an utmost significance to study the benefits of a diverse classroom in the manner that will provide students with sufficient knowledge about the importance of multiculturalism, but at the same time ease teacher's time spent at work. This paper examines the instances and benefits of diversity through the use of different strategies and analyses the multiculturalism of the 21st century merged in everyday classroom life.*

*Keywords: race, nation, culture, religion, multicultural education, strategies.*

INTRODUCTION

Diversity refers to the existence of a variety of cultural or ethnic groups within a single society. It is one of the most perceptible features of the time we live in and we encounter it in everyday life. Teachers should be conscious of a fact that each student in their classroom has an immense potential to be useful member of society and helpful to others—teachers, colleagues, and the community as a whole. Furthermore, diversity should be regarded as an encouragement of development and success, rather than an issue for students and educators.

A diverse classroom acts as a perfect environment for learning the multiple perspectives and diverse cultural patterns, and it prepares the students who spend their time in play with classmates from different backgrounds for the world outside the school. This paper examines diversity as one of the most perceptible features of our time, and presents a reader with the strategies for teachers and benefits that they, their students and society might attain through multicultural education.

MULTICULTURAL EDUCATION

Over the course of history, culture has been explained in many different ways. It has been defined as "the whole set of signs by which the members of a given society recognize…one another, while distinguishing them from people not belonging to that society" (UNESCO, 1992). Many scholars views culture as "the set of distinctive spiritual, material, intellectual and emotional features of a society or social group… (encompassing) in addition to art and literature, lifestyles, ways of living together, value systems, traditions and beliefs" (UNESCO, 2001).

Multicultural education refers to an idea that aims at promotion of educational equality and social justice. It carries multiple benefits and positively affects students moral and physical development. Diversity actively connects a variety of cultural, ethnic or linguistic groups within a single society. Teachers should be conscious of a fact that every student in their diverse classroom has a great potential to be a useful member of society and an invaluable resource for others. Furthermore, multicultural and multilingual classroom should be regarded as an encouragement of development and success, rather than an issue for students and educators.

Multicultural education refers to education and instructions which engage students of different cultures and linguistic backgrounds in the same activities, taking into consideration their views, beliefs and languages. It is designed to serve culturally different students in an equal and just way. On the other hand, intercultural education aims at promotion of understanding of differences amongst people and their cultures. Furthermore, it refers to teachings that shows a great respect and even promotes diversity in multiple areas of human life.

However, the major challenge which teachers might have when trying to promote the notion multiculturalism is dealing with rather natural and inherent tensions that arise when two or more different (or even opposing) world views meet. Those tensions, which reflect the diversity of co-existing values of multicultural world, usually can't be observed through an 'either/or' answer. However, the dynamic interaction between opposing views and aspects is what enriches education and multiculturalism.

This method of teaching and learning is based upon "consensus building, respect, and fostering cultural pluralism within societies" (Bennett, 1995). In conclusion, multicultural education intends to promote and incorporate positive cultural features and dialogue into language classroom atmosphere through creating a welcoming environment for everyone.

## THE POSITION OF A TEACHER

A language teacher as a leader of a classroom has the main role in managing the way in which teaching and overall interaction between the teacher and the students occur. From Zeichner's (1992) review of successful teaching approaches we can observe several key elements for effective teaching of culturally and linguistically diverse children:

1. Teachers should be completely aware of their own ethnic, cultural and linguistic identities.
2. Teachers emphasize high expectations for all students and believe that all of them are capable of making progress and success.
3. Teachers should develop good relations with their students and should not see them as "the others".
4. Teachers influence the creation of curricula that emphasize the development of higher-level cognitive and language skills.
5. Teachers influence the creation of curricula that include the contribution and perspectives of the different national, ethnic or cultural groups which compose the classroom.

In addition to the stated approaches, language teachers are to foster a good relationship with the families of their students, and they are to get to know different customs related to families' cultures. In conclusion, the real role of education and teachers is also stated in the Universal Declaration of Human Rights as following: "Education shall be directed to the full development of human personality and to the strengthening of respect for human rights and fundamental freedoms. It shall promote understanding, tolerance and friendship among all nations, racial and religious groups, and shall further the activities of the United Nations for the maintenance of peace" (Universal Declaration of Human Rights, 1948).

Furthermore, culturally responsive teaching is a term used nowadays for describing differentiated instructions and fitting teaching to the students' needs. Instructor Gay views this way of teaching as "using the cultural characteristics, experiences, and perspectives of ethnically diverse students as conduits for teaching them more effectively" (Gay, 2002 p. 106). She states that content and skills that are planned to be taught through educational systems are learned more easily when they belong to students' frames of reference and potential experiences (Gay, 2002).

## STRATEGIES FOR A CULTURALLY DIVERSE CLASSROOM

Strategies represent the ways through which a language teacher can make classroom a place of joy and quality learning. It is this welcoming environment that should push students to learn and behave in a good manner.

1. *Set and maintain high expectations for everyone regardless of their ethnicity, cutlre or language*

It has been proven that students whose teachers demonstrated high expectations for them learn better. Teachers who encourage students to identify and solve problems, and involve them in collaborative activities make their students aware of their ability to complete different tasks (Burris & Welner, 2005).

2. *Demonstrate care by learning about your students' needs, concerns and strengths*

Students show greater interest to participate in classroom activities when a teacher demonstrates care for them and their needs, hopes and dreams. Nel Noddings (1995) claims that "we should care more genuinely for our children and teach them to care" (p. 24).

3. *Learn about students' cultures and languages to better understand how and why they behave in certain ways in and out of the classroom.*

Teachers need to understand many different ways in which parents or care-givers might express concern about the education of their children in respect to their culture and language. For example, Gibson (1983) reports that Punjabi immigrant parents in California believe it is only the teacher's task to educate and that they as parents should not be involved in school activities. Furthermore, they showed that they care a lot about their native language, and that they are very cautious with their children learning another foreign language. All of this is to be taken into consideration when preparing a strategy for teaching (Gibson, 1983)

4. *Promote and encourage participation of parents or care-givers in school activities.*

Parents are a child's first teachers, but they are not necessarily aware how much they influence their children's development. Teachers can enhance parents' participation by informing them about the importance of a bond between home environment and children's learning in school (Saravia-Shore, 1992). Communication is crucial in language acquisition and learning.

5. *Choose culturally relevant curricula that recognize, incorporate, and reflect students' heritage*

Students certainly feel encouraged and motivated to study when they see that a teacher knows about and admits the contributions that their own racial or ethnic groups made to the community. This allows students to practice their language and other skills in real-life situations. They also realize that teacher values and appreciate each child's background, which creates more welcoming environment in a classroom.

6. *Include the arts in the curriculum.*

One of the best ways to enhance students' is to engage them in arts activities which promote dialogue on important issues. Providing opportunities for students to express their ideas and beliefs enables them to master talents and enhance multiple intelligences (Gardner, 1983).

### BENEFITS OF MULTICULTURAL EDUCATION

There are several very important benefits of multicultural education that might be encountered during the education period, but also afterwards throughout one's life. In this paper, we will present and examine benefits of breaking stereotypes and prejudices, biculturalism, acceptance of others and acculturation (Banks, 1993).

*Breaking stereotypes and prejudices*

Stereotypes represent core beliefs about certain characteristics that are believed to be features of a particular group or community. Stereotypes are usually incorrect and discriminatory and they, therefore, carry negative consequences for the people they are applied to.

Furthermore, stereotype threat, as explained by Steele (1997) refers to "the event of a negative stereotype about a group to which one belongs becoming self-relevant, usually as a plausible interpretation for something one is doing, for an experience one is having, or for a situation one is in, that has relevance to one's self-definition" (p. 686). Taking into consideration a danger of stereotyping, one might come to the conclusion how great benefit breaking stereotypes in a multicultural language classroom is.

*Biculturalism*

One of the greatest advantages of multicultural classroom is a formation of a bicultural perspective amongst students. Buriel et al. (1998) explained how knowledge of two cultures and languages allows students to better adjust to dual cultural need, and "may provide bicultural and bilingual students with more problem-solving strategies, interpersonal skills, and self-confidence for accessing academic resources at school and in their communities" (p. 294).

Therefore, students who become bicultural and bilingual can overcome communicational difficulties easier and are consequently able to co-exist and co-work in a culturally diverse environment.

*Acceptance of others*

In the twenty first century when humankind seeks toleration and acceptance in order to function well, the benefit of assimilation which is provided by multicultural education is of an utmost significance. In the second-language acquisition literature the mentioned assimilation is explained to be "the replacement of one's native culture, including language, values, social competencies and sense of identity, with that of another culture" (LaFromboise, Coleman, & Gerton, 1993). Therefore, the ultimate aim of assimilation might be considered to be social acceptance by members of the dominant group, which can be achieved in a classroom with culturally different beliefs and perspectives (LaFromboise et al., 1993).

*Acculturation*

The other benefit of a culturally diverse classroom is acculturation which refers alterations which result from a continuous contact between different cultures and languages: The mentioned changes come from learning the language, tradition and overall practices of the new culture (Sam & Berry, 2010).

However, the mentioned phenomenon, like assimilation, might result in the negligence of one's own culture (Buriel, 1993). As stated by Sam and Berry (2010), "individuals need to belong to a group in order to secure a firm sense of well-being." (p.475). When this is accomplished, students of a multicultural classroom are fully prepared to regard other cultures not only as equally important, but in a sense as their own.

## EDUCATION SYSTEMS IN BOSNIA AND HERZEGOVINA

Even though educational systems (plural form is used since B&H recognizes three different systems of education which are connected to three major nationalities) is regulated by law, Bosnia and Herzegovina often has many issues with implementing its unique education policies. One of the dilemmas that the government has is the question of whether to separate the public education system into special nation-oriented schools with different curricula (since the national diversity represents an obstacle instead of being beneficial). Many scientific papers have been written on this topic in journals in B&H but still no solutions have been offered that would be acceptable to all sides. Furthermore, this might be considered a political abuse of educational rights at the national level that avoids each attempt toward sincere democratization and acceptance of national, linguistic and religious diversity in the education systems.

There has been much manipulation of educational systems for political and ideological reasons from the time the war was over up until now. Education has often misused teaching the students different interpretations of the same events. For example, history textbooks in B&H may explain the initiation of the war as aggression, or a fight for liberation and national emancipation.

Moreover, sometimes educational systems divide students based on their nationality, language, and religion. Also, they tend to argue over the quality and acceptability of history and language textbooks. In these environments newly-formed education policies primarily reflect the superiority of majorities over minorities. In some cases minority children are allowed access only to education organized to serve the needs of the majority students.

This type of behaviour towards minority students demonstrates an unwillingness to accept Bosnian and Herzegovinian diversity present in the community. Although this exclusion goes against law, nationalist leaders are powerful enough to implement their will in schools, often making a phenomenon called "two schools under one roof". However, these problems occur infrequently in large cities such as Sarajevo, Tuzla, Bihać and Zenica, because they kept their multi-ethnic views and nurtured them through the educational system even during the war period (Pašalić-Krešo, 1999).

## CONCLUSION

The twenty first century is a century of an encounter of different cultures, beliefs and languages, and as such it exposes individuals to living in a culturally diverse society. Furthermore, educational institutions are becoming places of perspectives intertwining and are beginning to embrace the idea of the multiple cultures influencing students' development. In order for students to enhance their learning process and teachers to create a warm and welcoming atmosphere in their classes, it is of an utmost importance to analyze possible methods and approaches which can be used in everyday work at school. Multiculturalism has become one of the most intricate phenomena of the time in which we live. As a result of the different communities living together, there is a great demand for analyzing the benefits of a multicultural education in order to assure those who are suspicious of it, and believe that bringing students together can endanger their individual cultural customs and beliefs. In conclusion, this paper provides a reader with the information regarding the phenomenon of culturally diverse classroom, teachers' position and possible strategies for easing both educators' and students' work, defines multiculturalism in its essence and analyses four major benefits of multicultural education.

features that characterize a society or social group. It includes not only the arts and letters, but also modes of life, the fundamental rights of the human being, value systems, traditions and beliefs."